\documentclass[aps,preprintnumbers,11pt,amsmath,amssymb,nofootinbib]{revtex4}

\usepackage{braket}
\usepackage{float}
\usepackage{graphics,epsfig}
\usepackage{bm}
\usepackage{slashed}
\usepackage{graphicx}
\usepackage{amsmath}
\usepackage{amsfonts}
\usepackage{amssymb}
\usepackage{color}%
\usepackage{dcolumn}
\usepackage[
            pdfstartview=FitH,
            bookmarksnumbered=true,
            bookmarksopen=true,
            colorlinks,
            linkcolor=blue,
            anchorcolor=green,
            citecolor=red
            ]{hyperref}
\usepackage{enumerate}
\providecommand{\U}[1]{\protect\rule{.1in}{.1in}}

\def\A0{A^{(0)}}

\begin{document}
\baselineskip=0.6 cm \title{Quantum information metric of conical defect}

\author{Chong-Bin Chen$^{1,2}$}
\thanks{E-mail address: cchongb23@gmail.com}
\author{Wen-Cong Gan$^{1,2}$}
\thanks{E-mail address: ganwencong@gmail.com}
\author{Fu-Wen Shu$^{1,2}$}
\thanks{E-mail address: shufuwen@ncu.edu.cn}
\author{Bo Xiong$^{1}$}
\thanks{E-mail address: stevenxiongbo@gmail.com}
\affiliation{
$^{1}$Department of Physics, Nanchang University, Nanchang, 330031, China\\
$^{2}$Center for Relativistic Astrophysics and High Energy Physics, Nanchang University, Nanchang 330031, China}

\vspace*{0.2cm}
\begin{abstract}
\baselineskip=0.6 cm
\begin{center}
{\bf Abstract}
\end{center}
A concept of measuring the quantum distance between two different quantum states which is called quantum information metric is presented. The holographic principle (AdS/CFT) suggests that the quantum information metric $G_{\lambda\lambda}$ between perturbed state and unperturbed state in field theory has a dual description in the classical gravity. In this work we calculate the quantum information metric of a theory which is dual to a conical defect geometry and we show that it is $n$ times the one of its covering space. We also give a holographic check for our result in the gravity side. Meanwhile, it was argued that $G_{\lambda\lambda}$ is dual to a codimension-one surface in spacetime  and satisfies $G_{\lambda\lambda}=n_{d}\cdot\mbox{Vol}(\Sigma_{max})/L^{d}$. We show that the coefficient $n_d$ for conical defect should be rescaled by $n^2$ from the one for AdS. A limit case of conical defect --- the massless BTZ black hole--- is also considered. We show that the quantum information metric of the massless BTZ black hole disagrees with the one obtained by taking the vanishing temperature limit in BTZ black hole. This provides a new arena in differiating the different phases between BTZ spacetime and its massless cousin. 
\end{abstract}

\maketitle
\newpage
\vspace*{0.2cm}

%
\section{Introduction}
In the last two decades, our understanding of quantum gravity has been greatly enriched since the advent of the AdS/CFT correspondence conjecture \cite{Maldacena1998,Witten1998}. This conjecture has been expanded to gauge/gravity correspondence, a duality between more general spacetime background and quantum field theory. The holographic dictionary was established so that one can understand the physical quantities in gravity from its dual field theory equivalently, and vice versa. One of the most famous dictionaries in the past decade is the Ryu-Takayanagi (RT) formula \cite{Ryu:2006bv,Ryu:2006ef}. By computing the area of codimension-two extremal surface in the bulk, one can obtain the entanglement entropy in the quantum field on the boundary which is usually difficult to be calculated. Moreover, people begin to realize that quantum information theory plays crucial role in our understanding of gravity. More information theory quantities, like conditional entropy and mutual information have been introduced in discussion of gravitational issues.

One recent example is the proposal that the volume of an Einstein-Rosen bridge in spacetime is dual to the computational complexity of the corresponding quantum states of conformal field theory \cite{Susskind2014,Susskind2014_2,Stanford:2014jda}. This proposal derives a number of investigations and makes us to consider the dual of codimension-one volume. Moreover, a quantum information quantity called quantum information metric (QIM) which measures the distance between two infinitesimally different states,
\begin{equation}
|\braket{\Psi(\lambda+\delta \lambda)|\Psi(\lambda)}|=1-G_{\lambda\lambda } \delta \lambda^2+\mathcal{O}(\delta \lambda^3),
\end{equation}
was taken into account in the dictionary. $\ket{\Psi(\lambda)}$ is a one-parameter family state and the quantum information metric $G_{\lambda\lambda}$ is defined as minus the coefficient of the second order term of the absolute value of this overlap. In \cite{MIyaji:2015mia,Bak:2015jxd} the quantum information metric was argued to be dual to the volume of maximal spacelike surface
\begin{equation}\label{qim volume}
G_{\lambda\lambda}=n_{d}\cdot \frac{\mbox{Vol}(\Sigma_{max})}{L^{d}},
\end{equation}
where $n_d$ is a coefficient, $d$ is the spacetime dimension, $\Sigma$ is a spacelike surface and $L$ is the radius of the AdS$_{d+1}$ spacetime. This new dictionary may shed light on our understanding of codimension-one surface in holography (see for example \cite{Gan:2017qkz}).

In this paper we consider the quantum information metric of the dual CFT of conical defect geometry in three dimension. In bulk side, conical defect spacetime is a solution of the Einstein equation with static particle source at origin \cite{Deser:1983tn,Deser:1983nh}. It can be constructed by angular identification $\theta\sim\theta+2\pi/n$ on a pure AdS$_3$ spacetime. The conical defect can be regarded as a particular excited state so it is different from the vacuum AdS space. In the pure AdS space, it is possible to use the RT surfaces to probe all of the spacetime region if we consider the largest one, which is associated to the entanglement of half of the CFT on the boundary. However, when we consider the conical defect which is away from pure AdS, a so-called entanglement shadow that cannot be probed by minimal surfaces, will appear \cite{Balasubramanian:2014sra,Balasubramanian:2016xho}. To recover the physics in the entanglement shadows one should take into account the ``internal'' degrees of freedom of the dual CFT which is not spatially organized. In this way we give a computation of the quantum information metric for such conical defect dual CFT through the correlation function in the field theory. We find the quantum information metric of a conical defect is $n $ times the one of its covering space, which is pure AdS$_3$. Our calculation suggests that this result is related to the fact that the central charge of the conical defect is increased by a factor of $n$: $c=n\tilde{c}$. To verify our result, we also perform a holographic computation in the bulk. To proceed, we notice that the operator in CFT is dual to a local field in the bulk: $\Phi(z, x)\leftrightarrow \mathcal{O}(x)$. And the holographic relation $Z_{CFT}=Z_{bulk}$ implies one can turn to the computation of information metric in gravity side \cite{Trivella:2016brw}. Our results suggest that when we add a scalar field perturbation to the spacetime background, all scalar fields in copy of images should be considered. It gives the same result as the CFT computation. We also compare our result with (\ref{qim volume}) in conical defect and its covering space. We find that for conical defect this proposal also holds, but the coefficient $n_d^{defect}$ is rescaled by a factor $n^2$, i.e., $n^{defect}_{2}=n^2\cdot n^{AdS}_{2}$. We argue that this factor comes from the $\mathbb{Z}_n$ quotient on the original space, which is independent of dimension $d$. We also discuss the case $n\to\infty$, whose metric is a massless BTZ black hole. The massless BTZ has vanishing horizon so that one can approach the singularity by infinitely winding geodesics. We find the quantum information metric of masslass BTZ is inconsistent with the result which is obtained by taking limit $\beta\to\infty$ in information metric of BTZ. 

In section \ref{CD review} we review the conical defect geometry and its CFT dual. We show the Virasoro algebra and correlation function of operators $\mathcal{O}$ of the conical defect dual CFT. In section \ref{CD qim} we present the calculation of quantum information metric by path integral in field theory. We use this formulation to obtain the quantum information metric of the conical defect dual theory. We also recover the result from holographic point of view in dual gravity side and then compare this result with the proposal (\ref{qim volume}). In section \ref{MBTZ} we consider a limit case of the conical defect, which is the massless BTZ black hole case and give some remarks on it. Conclusions  are made in the last section.
\section{Conical defect geometry and its CFT dual}\label{CD review}
The conical defect geometry AdS$_3/\mathbb{Z}_n$ can be obtained by an angular identification of the global AdS$_3$ geometry. Its metric is the same as the global AdS$_3$ except the periodicity of the angular coordinate $\theta$ is $2\pi/n$ rather than $2\pi$\footnote{In general, $n$ need not be an integer, but the subalgebra of asymptotic symmetries only has the Virasoro form if $n\in \mathbb{Z}_+$ \cite{deBoer:2010ac}. We only consider integer $n$ in this paper.} , i.e. $\theta\sim\theta+2\pi/n$,
\begin{equation}
ds^2 = - \left( 1 + \frac{r^2}{L^2}\right) dt^2 + \left( 1 + \frac{r^2}{L^2}\right)^{-1} dr^2 + r^2 d\theta^2.
\label{covermetric}
\end{equation}
This is a solution of the negative constant-curvature Einstein equations with a static particle source placed at $r=0$ \cite{Deser:1983tn,Deser:1983nh}. Identifying $\theta\sim\theta+2\pi/n$ makes the space be a cone then the metric is not well defined at its vertex, where $r=0$. It is useful to change the coordinate from (\ref{covermetric}) by a rescaling
\begin{equation}
\theta' = n \theta \qquad {\rm and} \qquad r' = r / n \qquad {\rm and} \qquad t' = n t,
\label{rescalings}
\end{equation}
which results in
\begin{equation}
ds^2 = - \left( \frac{1}{n^2} + \frac{r'^2}{L^2}\right) dt'^2 + \left( \frac{1}{n^2} + \frac{r'^2}{L^2}\right)^{-1} dr'^2 + r'^2 d\theta'^2.
\label{defmetric}
\end{equation}
Now the periodicity of new angular coordinate is $2\pi$, i.e. $\theta'\sim\theta'+2\pi$, and there is still a singularity on $r'=0$. From these we can regard the pure AdS$_3$ as the covering space of the conical defect space. We call (\ref{covermetric}) as covering coordinates and (\ref{defmetric}) as defect coordinates. We have to 
point out that if we consider a cut-off of the infinite boundary $r=1/\epsilon$, we need also rescale the cut-off after this change of coordinates $\epsilon'=n\epsilon$, where $1/\epsilon$ and $1/\epsilon'$ are the IR cut-off in covering coordinates and defect coordinates respectively.

The field theory dual to the conical defect has been studied for a number of years \cite{deBoer:2010ac,Balasubramanian:2005qu,Balasubramanian:1999zv}. One approach is regarding the conical defect as an excited state of pure AdS$_3$ by adding a static particle on the origin \cite{Deser:1983nh,Balasubramanian:1999zv}. In the CFT picture the vacuum state dual to the vacuum AdS$_3$ and the conical defect is a particular excited state.

According to the Brown-Henneaux method in constructing the asymptotic symmetry algebra of the AdS$_3$, the Virasoro algebra which preserves the diffeomorphisms of the covering space AdS$_3$ is \cite{Brown:1986nw}
\begin{equation}
[\tilde{L}_k,\tilde{L}_s] = (k-s) \tilde{L}_{k+s} + \frac{\tilde{c}}{12} k(k^2 - 1) \delta_{k+s} ,
\end{equation}
where $\tilde{c}$ is the central charge of the covering theory\footnote{In this paper, all quantities associated with the covering space are marked by a tilde to differentiate them from those belonging to the conical defect.}. But in conical defect theory, not all diffeomorphisms will descend to defect theory because of the restriction of $\mathbb{Z}_n$ symmetry. The Virasoro algebra of the defect theory is a subalgebra generated by $\tilde{L}_{nk}$ \cite{Balasubramanian:2014sra,deBoer:2010ac}:
\begin{equation} \label{virasoro}
L_k = \frac{1}{n} \tilde{L}_{nk}, \ \ k \neq 0 ~~~~~~~~~~~~~ L_0 = \frac{1}{n} \left(\tilde{L}_0 - \frac{\tilde{c}}
{24} \right) + \frac{n\tilde{c}}{24}
\end{equation}
The defect theory has the central charge $c=n\tilde{c}$. Since the central charge is related to the AdS$_3$ radius and gravitational constant as $c=3L/2G$, this indeed corresponds to the rescaling of the gravitational constant $\tilde{G}=nG$ in the bulk.

In the CFT, there are some ``internal'' degrees of freedom which are not spatially organized. The question is how these degrees of freedom entangle and how to describe them. The internal degrees of freedom are usually gauged. In \cite{Balasubramanian:2014sra} the authors defined a so-called ``entwinement'' to probe these gauged degrees of freedom. They deal with this problem in this way: first embedding the theory in a larger theory (the covering space) where the degrees of freedom are not gauged. After ``ungaging'' the gauge symmetry we compute the $\mathbb{Z}_n$-invariant quantities which is the sum of quantities of all gauged copies in the ungauged theory. For example, we should sum over images of correlation to compute the the correlation function of operator $\mathcal{O}$ in dual defect CFT$_{c}$. The operator $\mathcal{O}$ is given by \cite{Balasubramanian:2014sra,Arefeva:2016wek}:
\begin{equation}
\mathcal{O}(\tau, \theta) = \sum_{k=0}^{n-1} g^k\tilde{\mathcal{O}}(\tau, \theta) =\sum_{k=0}^{n-1} e^{i \frac{2\pi k}{n} \frac{\partial}{\partial \theta}} \tilde{\mathcal{O}}(\tau, \theta), \label{operator}
\end{equation}
where $g^k$ is a $\mathbb{Z}_n$ generator. So one can use the operator $\mathcal{\tilde{O}}$ in covering theory CFT$_{\tilde{c}}$ to compute the two-point function in defect theory CFT$_c$ as
\begin{eqnarray}
\left< \mathcal{O}(\tau_1, \theta_1) \mathcal{O}(\tau_2, \theta_2) \right>
&=& \sum_{a=0}^{n-1} \sum_{b=0}^{n-1}  e^{i \frac{2\pi a}{n} \frac{\partial}{\partial \theta_1}}  e^{i \frac{2\pi b}{n} \frac{\partial}{\partial \theta_2}} \langle \tilde{\mathcal{O}}(\tau_1, \theta_1) \tilde{\mathcal{O}}(\tau_2, \theta_2) \rangle \nonumber\\
&=& \sum_{a=0}^{n-1} \sum_{b=0}^{n-1}   \langle \tilde{\mathcal{O}}\big(\tau_1, \theta_1+ \frac{2\pi a}{n} \big) \tilde{\mathcal{O}}\big(\tau_2, \theta_2+\frac{2\pi b}{n}\big) \rangle \label{cd 2-p},
\end{eqnarray}
where we have obtained the expression for the correlation function as a sum over all images.

The correlation function in the conical defect can be correctly constructed from GKPW and geodesic approximation, up to a factor \cite{Arefeva:2016wek,Ageev:2015qbz,Arefeva:2015zra}. There is also a so-called ``entwinement'' which discribes the internal degrees of freedom by these long geodesics as formed in \cite{Balasubramanian:2014sra}. These non-minimal geodesics can probe the entanglement shadow in the spacetimes, which is the region the minimal one cannot reach in the conical defect geomentry. The same method of images for computing Green functions was also used in BTZ spacetimes \cite{Balasubramanian:2005qu,KeskiVakkuri:1998nw,Lifschytz:1993eb,
BDHM:1998}. In the next section we will use (\ref{cd 2-p}) to compute the correlation function in the dual boundary theory of the conical defect(CFT$_{c}$) and to see the relation with the one in the dual boundary theory of the  covering space(CFT$_{\tilde{c}}$).

\section{Quantum information metric for conical defect}\label{CD qim}
\subsection{CFT computation}
In this subsection we try to calculate the quantum information metric of the CFT dual to the conical defect geometry from the boundary theory side. We first clarify the analysis in \cite{MIyaji:2015mia,Bak:2015jxd} for the general $d$-dimensional CFT and then turn to discuss the conical defect case. More details are in Appendix \ref{review qim}.

We consider a $d$-dimensional unperturbed CFT whose Euclidean Lagrangian is $\mathcal{L}_0$ and the corresponding ground state is $\ket{\Psi_0}=\ket{\Psi(\tilde{\lambda}=0)}$, where $\tilde\lambda$ is the parameter of this state. Now one can add a little deformation to this original CFT by adding $\delta \mathcal{L}_0=\delta\tilde{\lambda}\tilde{\mathcal{O}}$ to $\mathcal{L}_0$, so the perturbed Lagrangian can be denoted by $\mathcal{L}_1=\mathcal{L}_0+\delta \mathcal{L}_0$. and the corresponding ground state of perturbed field is denoted by $\ket{\Psi_1}=\ket{\Psi(\tilde{\lambda}+\delta\tilde{\lambda})}$. Difference between the unperturbed state and the perturbed one is given by
\begin{equation}\label{overlap2.1}
\braket{\Psi_1 (\epsilon)|\Psi_0}=\frac{\braket{\exp\left( -\int_{\epsilon}^{\infty} d \tau \int d^{d-1}x \delta \tilde{\lambda} \tilde{\mathcal{O}}(\tau, x)\right)}}{\braket{\exp\left( -(\int_{-\infty}^{-\epsilon}+\int_{\epsilon}^{\infty}) d \tau \int d^{d-1}x \delta \tilde{\lambda} \tilde{\mathcal{O}}(\tau, x)\right)}^{1/2}},
\end{equation}
where $\ket{\Psi_1 (\epsilon)}$ is the regularized ground state to which we introduce a cut-off $\epsilon$ at $\tau=0$ because there is a discontinuity at this location, see (\ref{reg state}) in Appendix \ref{review qim}. Quantum information metric is defined by minus the coefficient of the second order term of the absolute value of this overlap
\begin{equation}
|\braket{\Psi_1 (\epsilon)|\Psi_0}|=1-G_{\tilde{\lambda} \tilde{\lambda}} \delta \tilde{\lambda}^2+\tilde{\mathcal{O}}(\delta \tilde{\lambda}^3),
\end{equation}
It's not difficult to obtain its expression from (\ref{overlap2.1}):
\begin{equation}
G_{\tilde{\lambda} \tilde{\lambda}} =\frac{1}{2}\int d^{d-1}x_1 \int d^{d-1}x_2\int_{-\infty}^{-\epsilon} d \tau_1 \int_{\epsilon}^{\infty}d \tau_2\braket{\tilde{\mathcal{O}}(\tau_1,x_1)\tilde{\mathcal{O}}(\tau_2,x_2)} \label{2-p},
\end{equation}
where we have assumed the one-point function is zero and used the reversal symmetry relation of two-point function. One can also consider the CFT which lives on a cylinder $\mathbb{R}\times S^{d-1}$. The quantum information metric of this case is more non-trivial because it has the universal term in general. The derivation of the quantum information metric of cylinder is given by \cite{Bak:2017rpp}
\begin{equation}
G_{\tilde{\lambda} \tilde{\lambda}} =\frac{1}{2}\int d^{d-1}\Omega_1 \sqrt{g_{S^{d-1}}}\int d^{d-1}\Omega_2 \sqrt{g_{S^{d-1}}}\int_{-\infty}^{-\epsilon} d \tau_1 \int_{\epsilon}^{\infty}d \tau_2\braket{\tilde{\mathcal{O}}(\tau_1,\Omega_1)\tilde{\mathcal{O}}(\tau_2,\Omega_2)},
\end{equation}
where $g_{S^{d-1}}$ is the determinant of the metric of the sphere and it cannot be ignored for $d>2$. While for $d=2$, it reduces to $1$. In what follows we only cosider this case.

Now let us turn to compute the quantum information metric of the field theory which is dual to the conical defect geometry. Before doing that, we should be careful with the normalization constant of the correlation functions. For a CFT which lives on a cylinder, we have\cite{Bak:2015jxd,Bak:2017rpp}:
\begin{equation}
\braket{\tilde{\mathcal{O}}(\tau_1,\theta_1)\tilde{\mathcal{O}}(\tau_1,\theta_2)}=\frac{{\cal N}_{\Delta,\tilde{\kappa}}}{(2 \cosh(\tau_1-\tau_2)-2 \cos(\theta_1- \theta_2))^{\Delta}},
\end{equation}
In order to guarantee consistence with the bulk computation, one choose the following normalization constant:
\begin{equation}\label{normalization}
{\cal N}_{\Delta,\tilde{\kappa}}= \frac{ L^{d-1}\, d \, \Gamma(\Delta ) }{\tilde{\kappa}^2\pi^{\frac{d }{2}} \Gamma(\Delta -\frac{d}{2})  }
\end{equation}
which is related to the Newton constant ${\cal N}_{\Delta,\tilde{\kappa}} \sim 1/\tilde{\kappa}^2=1/8\pi \tilde{G}$ (one can find the derivation of this constant in Appendix \ref{normalizationconstant}). It was mentioned in the last section that the gravitational constant for conical defect and covering space is different by a rescaling due to (\ref{virasoro}). In other words, when we represent quantities of the conical defect in terms of the ones of the covering space, this constant should also be changed. As a consequence,  this will give us a rescaling $\tilde{\kappa}^2=n\kappa^2$ .

Now let us see the quantum information metric of the conical defect, by making use of the correlation function of the conical defect (\ref{cd 2-p}) and the definition of the quantum information metric (\ref{2-p})
\begin{eqnarray}
G^{defect}_{\lambda \lambda}
&=&\frac{1}{2}\int_{0}^{\frac{2\pi}{n}} d\theta_1 \int_{0}^{\frac{2\pi}{n}} d\theta_2\int_{-\infty}^{-\epsilon} d \tau_1 \int_{\epsilon}^{\infty}d \tau_2\braket{\mathcal{O}(\tau_1,\theta_1)\mathcal{O}(\tau_2,\theta_2)}_{\kappa}\nonumber\\
&=& \frac{1}{2}\sum_{a=0}^{n-1} \sum_{b=0}^{n-1}\int_{0}^{\frac{2\pi}{n}} d\theta_1 \int_{0}^{\frac{2\pi}{n}} d\theta_2\int_{-\infty}^{-\epsilon} d \tau_1 \int_{\epsilon}^{\infty}d \tau_2\langle \tilde{\mathcal{O}}\big(t_1, \theta_1+ \frac{2\pi a}{n} \big) \tilde{\mathcal{O}}\big(t_2, \theta_2+\frac{2\pi b}{n}\big) \rangle_{\kappa}. \label{cd QIM}
\end{eqnarray}
The subscripts $\kappa$ means the normalization constant of the correlation functions is related to Newton constant $\kappa^2 \sim G$ of the conical defect geometry. Our aim is to find the relation of the quantum information metric between the conical defect $G^{defect}_{\lambda \lambda}$ and the covering space $G^{AdS}_{\tilde{\lambda} \tilde{\lambda}}$. Using the identity
\begin{eqnarray}\label{identity}
\sum_{a=0}^{n-1}\int_{0}^{\frac{2\pi}{n}} d\theta f(\theta+ \frac{2\pi a}{n})
&=&\sum_{a=0}^{n-1}\int_{\frac{2\pi a}{n}}^{\frac{2\pi(a+1)}{n}}dx_a f(x_a)\nonumber\\
&=&\int_{0}^{2\pi}dx f(x).
\end{eqnarray}
where $x_a=\theta+ 2\pi a/n$ and $x_b=\theta+ 2\pi b/n$, we can rewrite (\ref{cd QIM}) as
\begin{eqnarray}
G^{defect}_{\lambda \lambda}
&=&\frac{1}{2}\int_{0}^{2\pi} dx_1 \int_{0}^{2\pi} dx_2\int_{-\infty}^{-\epsilon} d \tau_1 \int_{\epsilon}^{\infty}d \tau_2\braket{\tilde{\mathcal{O}}(\tau_1,x_1)\tilde{\mathcal{O}}(\tau_2,x_2)}_{\kappa}\nonumber\\
&=&\frac{n}{2}\int_{0}^{2\pi} dx_1 \int_{0}^{2\pi} dx_2\int_{-\infty}^{-\epsilon} d \tau_1 \int_{\epsilon}^{\infty}d \tau_2\braket{\tilde{\mathcal{O}}(\tau_1,x_1)\tilde{\mathcal{O}}(\tau_2,x_2)}_{\tilde{\kappa}}.
\end{eqnarray}
In the second line we have changed the Newton constant from $\kappa$ to $\tilde{\kappa}$ such that it introduces a rescaling factor $n$. We note that the integral multiplied by $1/2$ in the second equality is nothing but the quantum information metric of the covering space, so we have
\begin{equation}\label{main result}
G^{defect}_{\lambda \lambda}=n\cdot G^{AdS}_{\tilde{\lambda} \tilde{\lambda}}.
\end{equation}
From this relation we conclude that the quantum information metric of conical defect is larger than the pure AdS case in general. The difference comes from the Virasoro algebra of the dual CFT of the conical defect. Specifically, $L_0$ has a fractionated spectrum (\ref{virasoro}), which results in the different central charges between two theories.

We are going to give more remarks on this result now. As we claimed above, the theory dual to conical defect has a fractionated spectrum, which comes from the fact that the spectrum of $L_0$ is rescaled by $1/n$ relative to $\tilde{L}_0$. That means the theory which is dual to the conical defect has smaller spectral gap so more ``internal'' degrees of freedom than the one dual to the covering space. One can regard their relation as encapsulating $n$ degrees of freedom of conical defect into a single degree of freedom of covering space, which leads to $\tilde{c}=c/n$ \cite{Balasubramanian:2014sra}, because the central charge measures the degree of freedom of a quantum field. In other words, the dual quantum system of the conical defect is more complex than the covering one. On the other hand, the quantum information metric also has the physical meaning as the fidelity susceptibility of the quantum system $\mathcal{F}(\lambda, \lambda+\delta\lambda)=1-G_{\lambda \lambda} \delta \lambda^2+\mathcal{O}(\delta \lambda^3)$, which measures the distance between a perturbed ground state and its unperturbed ground state. So we can expect that as a marginal deformation with the same amplitude is added to the state dual to the conical defect and the covering space respectively, more perturbative changes of the state will be observed for the conical defect comparing to the covering, because of the greater number of degrees of freedom and correlations. This results in less fidelity $\mathcal{F}^{defect}(\lambda, \lambda+\delta\lambda)\le\mathcal{F}^{AdS}(\tilde{\lambda}, \tilde{\lambda}+\delta\tilde{\lambda})$, and is consistent with \eqref{main result} where $G^{defect}_{\lambda \lambda}\ge G^{AdS}_{\tilde{\lambda} \tilde{\lambda}}$.

\subsection{Bulk computation}
In this subsection we perform the holographic set up for an explicit checks to our CFT calculation. We will compute the quantum information metric for an exactly marginal deformation in gravity side by using the holographic method. Here we just use some main result from \cite{Trivella:2016brw} and more details can be found in Appendix \ref{review qim}.

Let $Z_2$ be the partition function of a theory that only deforms the CFT for $\tau>0$, i.e.
\begin{equation}
Z_2=\int \mathcal{D} \varphi \exp \left(-\int_{-\infty}^{0} d \tau \int d^{d-1}x  \mathcal{L}_{0}- \int_{0}^{\infty} d \tau \int d^{d-1}x ( \mathcal{L}_0 +\delta \tilde{\lambda} \tilde{\mathcal{O}})\right)
\end{equation}
One can write (\ref{overlap-pi}) as
\begin{equation}\label{bulk overlap}
\braket{\Psi_1|\Psi_0}=\frac{Z_2}{(Z_1 Z_{0})^{1/2}},
\end{equation}
where $Z_0$ is the partition function of the unperturbed theory and $Z_1$ is the partition function of the perturbed theory for $-\infty<\tau<\infty$ (see (\ref{z0}) and (\ref{z1}) in appendix \ref{review qim}).

The holographic dual conjecture allows us to calculate the partition functions of the CFT from the gravity side. Particularly in the large $N$ limit we can write the partition functions as $Z_k=e^{-I_k}$, where $I_k$ is the on-shell action which corresponds to the field theories with $k=0, 1, 2$, respectively. After a deformation, the on-shell action is perturbed by adding a contribution of a scalar field probing on the AdS space, $I_k=I_{AdS}+\delta I_k$, with
\begin{eqnarray}\label{EQ2}
\delta I_k
&=&\frac{1}{2\tilde{\kappa}^2}\int_{\mathcal{M}} d^{d+1}x\sqrt{g_0}(g_0^{\mu\nu}\partial_{\mu}\tilde{\Phi}_k\partial_{\nu}\tilde{\Phi}_k+m^2\tilde{\Phi}_k^2)\nonumber\\
&=&\frac{1}{2 \tilde{\kappa}^2}\int_{\partial \mathcal{M}}\sqrt{\gamma_0} n_{\mu} g^{\mu \nu}_0 \tilde{\Phi}_k \partial_\nu \tilde{\Phi}_k,
\end{eqnarray}
where $g_0^{\mu\nu}$ is unperturbed AdS metric. In the second line we have used the equation of motion and have written $\delta{I_k}$ as boundary contribution by integrating by parts. $\tilde{\Phi}_k$ can be obtained by the boundary to bulk propagator in holography. Then one can write the overlap (\ref{bulk overlap}) as:
\begin{eqnarray}
\braket{\Psi_1|\Psi_0}&=&\frac{Z_{2}}{\sqrt{Z_1 Z_0}}=\exp \left(-I_{AdS}-\delta I_{2} + \frac{1}{2}(I_{AdS}+\delta I_{1}+I_{AdS})\right) \nonumber\\
&=& \exp \left( -\delta I_{2} + \frac{1}{2} \delta I_{1} \right). \label{action qim}
\end{eqnarray}
We now consider the quantum information metric for the vacuum state of a CFT living on a cylinder. The operator $ \tilde{\mathcal{O}} $ is dual to a probing scalar field in global AdS whose metric is given by
\begin{equation}\label{global}
ds^2=(1+r^2) d\tau^2+\frac{dr^2}{1+r^2}+r^2 d\Omega^2_{d-1}, 
\end{equation}
where we have set the AdS radius to one for simplicity.

We consider the marginal deformation, i.e. $\Delta=d$. There is only $\delta I_2$ contribution to the quantum information metric \cite{Trivella:2016brw}. So we have only the boundary condition of the scalar field $\Phi_2$: 
\begin{equation} \label{coupling const}
\lim_{r\rightarrow \infty}\tilde{\Phi}_2(r,\tau)=\begin{cases}
\delta \tilde{\lambda} & \text{if  }\tau>0 \\
0& \text{if  }\tau<0.
\end{cases}
\end{equation}
By adopting this boundary condition and solving the differential equation of $\tilde{\Phi}_2$, we can obtain the corresponding action contribution of probing scalar field $\delta I_2$ (\ref{action qim}).
\begin{figure}
\centering
\includegraphics[scale=0.8]{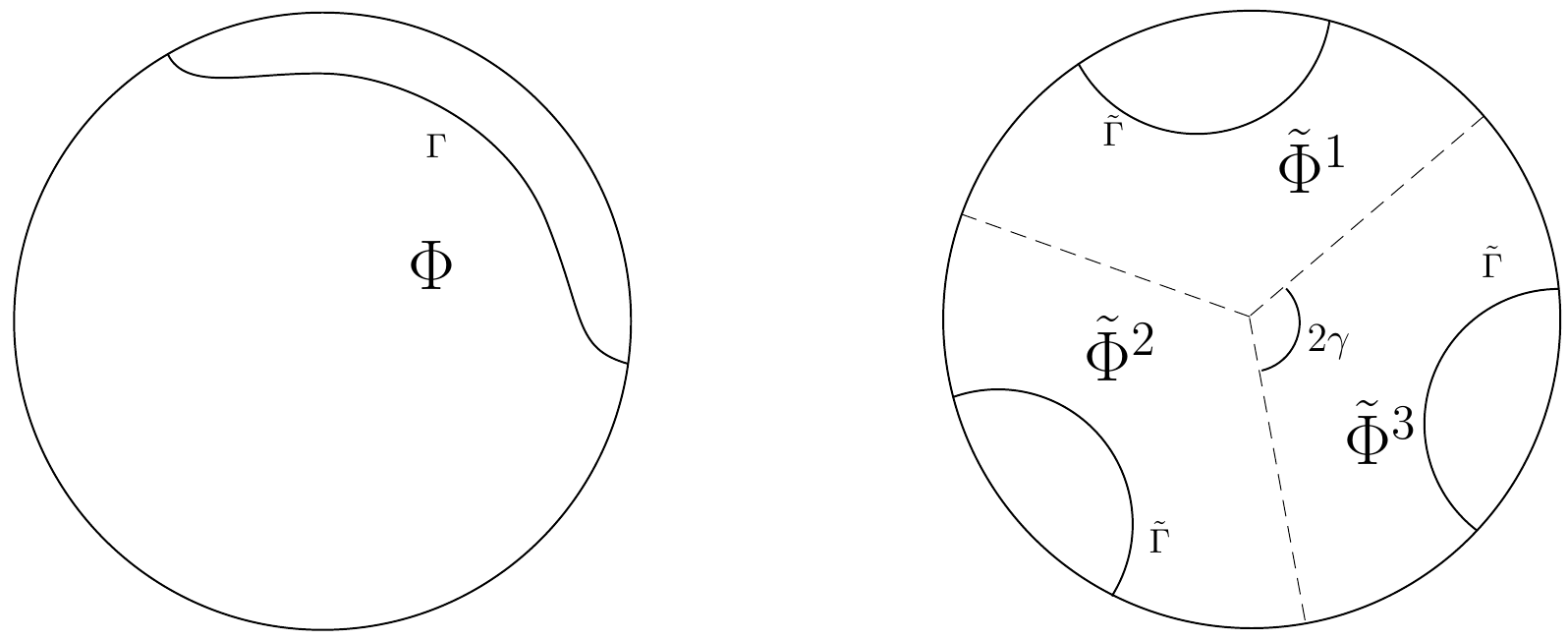}
\caption{(Left) Scalar field $\Phi$ on the conical defect geometry. $\Gamma$ is a minimal geodesic corresponds to a subregion of the system. It descends to a minimal geodesic $\tilde{\Gamma}$ on its covering space. (Right) The covering space of the conical defect. $\tilde{\Phi}^{i}$ are the scalar fields on it. $\Phi$ in the left is $\mathbb{Z}_n$ symmetric $\tilde{\Phi}$ mode which is a solution of equation in each conical defect copy. Let $\tilde{\Phi}^i$ denotes the symmetrized modes of AdS in $i$th single copy.}\label{scalarfield}
\end{figure}

We turn to the quantum information metric of conical defect. As shown above, for marginal perturbations, only $\delta I_2$ survives in (\ref{action qim}). By definition, we have $G_{\lambda \lambda}\delta \lambda^2=\delta I^{defect}_2$.
This implies that, to obtain the $G_{\lambda\lambda}$, we should find out the coupling constant $\delta\lambda$ and deformation $\delta I_2^{defect}$ for conical defect. To proceed, we notice that there is a relation between scalar fields in the conical defect and the covering $\Phi(r=\infty, \tau, \theta)=\tilde{\Phi}(r=\infty, \tau, \theta)$, and it can be understood as the following. The scalar wavefunctions in the conical defect or the covering space can be obtained by solving the Klein-Gordon equation $(\square-m^2)\Phi=0$ in the corresponding bulk geometry backgrounds. The solutions can be written as a usual ansatz \cite{Arefeva:2016wek,Cresswell:2017mbk}: $\Phi(r, t, \theta)=e^{i\omega t}Y_l(\theta)R(r)$. In covering space the angular dependent term $Y_l$ is solved as an eigenproblem and is given by
\begin{equation}
\tilde{Y_l}(\theta)=e^{il\theta},~~~~~~~~~~\tilde{Y_l}(\theta+2a\pi)=\tilde{Y_l}(\theta),~~~~~~~~~~l,a\in \mathbb{Z}.
\end{equation}
Similarly, the angular dependent term for conical defect can be written as the same form but with different periodicity \cite{Arefeva:2016wek}:
\begin{equation}
Y_j(\theta)=e^{inj\theta},~~~~~~~~~~Y_j(\theta+\frac{2b\pi}{n})=Y_j(\theta),~~~~~~~~~~j,b\in \mathbb{Z}.
\end{equation}
Therefore the $\Phi$ modes of the conical defect is a subset of the $\tilde{\Phi}$ modes with $l=nj$. In other words, the solution of the Klein-Gordon equation in the conical defect can be given by the $\mathbb{Z}_n$ symmetric $\tilde{\Phi}$ modes and restricting $\tilde{\Phi}$ to a single $\mathbb{Z}_n$ image, as showed in (FIG. \ref{scalarfield}). Hence near the boundary we have the same coupling constants in the two theories, i.e. $\delta \lambda = \delta \tilde{\lambda}$. In the rest of this paper we only use $\delta\lambda$ to denote the coupling constant, and will not differiate them from now on.

Next let us turn to the calculation of $\delta I^{defect}_2$. It is clear that to obtain the $\delta I^{defect}_2$, we simply need to sum over all contributions of the scalar fields probing the unperturbed background of all images in covering space:
\begin{eqnarray}
\delta I_2^{defect}
&=&\frac{1}{2 \kappa^2}\sum_{k=0}^{n-1}\int_{\partial \mathcal{M}_k}\sqrt{\gamma_0} n_{\mu} g^{\mu \nu}_0 \Phi_2(\theta + \frac{2\pi k}{n}) \partial_\nu \Phi_2(\theta + \frac{2\pi k}{n})\big{|}_{r=\infty} \nonumber\\
&=&\frac{n}{2 \tilde{\kappa}^2}\int_{\partial \tilde{\mathcal{M}}}\sqrt{\gamma_0} n_{\mu} g^{\mu \nu}_0 \tilde{\Phi}_2 \partial_\nu \tilde{\Phi}_2\big{|}_{r=\infty},
\end{eqnarray}
where $\partial \mathcal{M}_k$ and $\partial \tilde{\mathcal{M}}$ are the boundaries of the conical defect and the covering space respectively. In the second line we again have used the identity (\ref{identity}) to change the integral from conical defect $\partial \mathcal{M}$ to covering space $\partial \tilde{\mathcal{M}}$. We also change the gravitational constant by using $\tilde{\kappa}^2=n\kappa^2$. All these observations result in
\begin{equation}
\delta I_2^{defect}=n\cdot \delta I_2^{AdS},
\end{equation}
where $\delta I_2^{AdS}$ is the perturbation of the covering space, which is just the pure AdS space. Recalling $G_{\lambda\lambda}\delta\lambda^2=\delta I_2$, we immediately get $G^{defect}_{\lambda \lambda}=n\cdot G^{AdS}_{\lambda \lambda}$, as expected.

By introducing a cut-off $\epsilon$, we get the following expression for the quantum information metric of the global AdS (\ref{globalG}),
\begin{equation}
G^{defect}_{\lambda \lambda}=\frac{nL^{d-1} \Gamma\left(\frac{1+d}{2}\right)}{2 \tilde{\kappa}^2 \sqrt{\pi} \Gamma(d/2)} \cdot \frac{\text{Vol}(\mathbb S^{d-1})}{(d-1)\epsilon^{d-1}}+\cdots,
\end{equation}
which has a cut-off independent universal constant. For even $d$ the universal term is the $\mathcal{O}(\epsilon^0)$ and for odd $d$ it is the coefficient of a logarithmic term.

\subsection{Comparing with the QIM/Volume correspondence}
In \cite{MIyaji:2015mia} the authors argued that $G_{\lambda \lambda}$ of a $(d+1)$-dimension CFT deformed by a marginal perturbation can be holographically estimated by
\begin{equation}
G_{\lambda\lambda}=n_{d}\cdot \frac{\mbox{Vol}(\Sigma_{max})}{L^{d}},  \label{volume}
\end{equation}
where $n_d$ is an coefficient and $L$ is the AdS radius. $\mbox{Vol}(\Sigma_{max})$ is the maximal volume of a $d$-dimension spacelike surface in the bulk geometry, which is generally infinite and we should regularized with a cut-off $\epsilon$. Next we are going to utilize our result to discuss this holographic conjecture. We find that the relation (\ref{volume}) is preserved. However, the coefficient $n_d$ in conical defect is different from the one in covering space.

In the global AdS$_{d+1}$ case whose metric is given by (\ref{global}), the quantum information metric can be computed from the volume of a time slice. But we now prefer to use the coordinate (\ref{defmetric}) and we introduce the cut-off $\epsilon'=n\epsilon$. When $n=1$ it is the pure AdS (we have set the AdS radius to one):
\begin{equation}
G^{AdS}_{\lambda\lambda}=n^{AdS}_dV'_{d-1}\int^{r'_{\infty}}_{0}\frac{r'^{d-1}}{\sqrt{r'^2+1}}dr' =\frac{n^{AdS}_dV'_{d-1}}{(d-1)\epsilon^{d-1}}+\cdots,
\end{equation}
where $V'_{d-1}=\text{Vol}(\mathbb S^{d-1})$ and $\epsilon=\epsilon'$ while $n_d^{AdS}$ is given by (\ref{nd}) with $L=1$. Particularly in $d=2$, we have
\begin{equation}
G^{AdS}_{\lambda\lambda}(d=2)=n^{AdS}_2V'_1\cdot\left(\frac{1}{\epsilon}-1+\frac{\epsilon}{2}-\frac{\epsilon^3}{8}+\cdots\right),
\end{equation}
which has a cut-off independent term.

We turn to estimate the quantum information metric of conical defect in $d=2$ by using the coordinate (\ref{defmetric}). We calculate the volume of the time slice for a conical defect geometry and we obtain
\begin{eqnarray}\label{2d df qim}
G^{defect}_{\lambda\lambda}&=&n^{defect}_2V_1\int^{r'_{\infty}}_{0}\frac{r'}{\sqrt{r'^2+\frac{1}{n^2}}}dr'\\
&=&n^{defect}_{2}V_1\cdot \left(\frac{1}{\epsilon'}-\frac{1}{n}+\frac{2\epsilon'}{2n^2}-\frac{\epsilon'^3}{8n^4}+\cdots\right),
\end{eqnarray}
where $n^{defect}_2$ is a coefficient in conical defect geometry. Replacing $\epsilon'$ by $n\epsilon$ and comparing with our result (\ref{main result}), we have $n^{defect}_{2}=n^2\cdot n^{AdS}_{2}$.

We observe that on the premise of generality of the argument (\ref{volume}), when we make a $\mathbb{Z}_n$ quotient of the original geometry, the coefficient $n_2$ should be rescaling by $n^2$. Generally speaking, in $d+1$-dimensional we have
\begin{equation}
n^{defect}_{d}=n^2\cdot n^{covering}_{d}
\end{equation}

\begin{figure}
\centering
\includegraphics[scale=0.8]{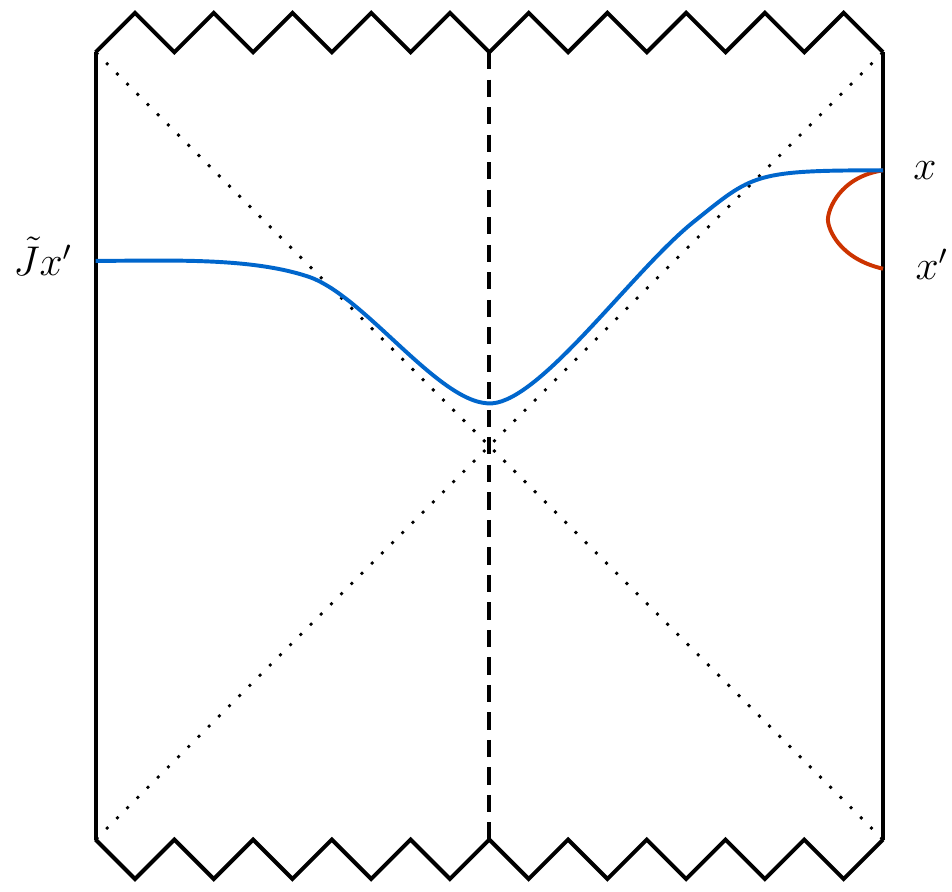}
\caption{The Penrose diagram of a BTZ black hole. The geon black hole can be obtained by a quotient of the BTZ, which is the vertical axis in the diagram. The two-point function includes two distinct contributions: one from correlations between two operators on the same boundary (red),  the other from correlations between two operators on different boundaries (blue).}\label{PenroseD}
\end{figure}
One more example comes from a geon black hole. It was shown in \cite{Sinamuli:2016rms} how to construct a geon black hole. The $\mathbb{RP}_2$ geon can be constructed by taking a quotient by the $\mathbb{Z}_2$ action on the BTZ black hole. The BTZ black hole metric is given by \cite{Banados:1992wn,Banados:1992gq}
\begin{equation}
ds^2 = -\left(\frac{r^2 - r_+^2}{L^2}\right) \,dt^2
+ \left(\frac{L^2}{r^2 - r_+^2}\right) dr^2 + r^2 d\theta^2.
\label{btzmetric}
\end{equation}
The map $\tilde{J}$ of $\mathbb{Z}_2$ quotient maps a point at $(t,\theta)$ in an asymptotically region to the one at $(-t,\theta+\pi)$ in another asymptotically region. The boundary manifold is a Klein bottle which is non-orientable. The $(d+1)$-dimensional AdS Schwarzchild planar black hole is analogous to the BTZ case. As a $\mathbb{Z}_2$ quotient of BTZ and AdS Schwarzchild planar black hole, our result implies
\begin{equation}\label{4nd}
n^{GEON}_2 = 4\cdot n^{BTZ}_2~~~~~~~~~~~~~n^{SAdSgeon}_d = 4\cdot n^{SAdS}_d,
\end{equation}
which agree with the result in \cite{Sinamuli:2016rms}. We can see this result is not related to the dimension $d$ because it comes from the quotient on the original manifold. They compute the correlation function in geon dual theory by summing over the two image contributions from the result of the quotient in BTZ \cite{Louko:2000tp, Guica:2014dfa}, i.e.
\begin{equation}
\braket{\mathcal{O}(x) \mathcal{O}(x')}_{\text{geon}}=\braket{ \mathcal{O}(x) \mathcal{O}(x') }_{\text{BTZ}}+\braket{ \mathcal{O}(x) \mathcal{O}(\tilde{J}x') }_{\text{BTZ}},
\end{equation}
where the first contribution comes from the BTZ case and the second contribution is the two-point function of two points in different boundary (FIG. \ref{PenroseD}). They found the second contribution to the quantum information metric is same as the first one. So the contribution is double of that of the original black hole, i.e. $G_{\lambda\lambda}^{\text{geon}}=2G_{\lambda\lambda}^{\text{BTZ}}$. On the other hand, the identification $(t, \theta)\sim(-t, \theta+\pi)$ shows a reflection around the vertical axis in Penrose diagram of the BTZ black hole, as shown in (FIG. \ref{PenroseD}). So the volume $\text{Vol}^{\text{geon}}(\Sigma)$ is reduced to one half of the BTZ volume, i.e. $\text{Vol}^{\text{geon}}(\Sigma)=\text{Vol}^{\text{BTZ}}(\Sigma)/2$. Combining with \eqref{volume}, it gives a factor $4$ on the coefficient (\ref{4nd}).

We can also use our method of holographic computation for the $\mathbb{RP}^2$ geon which is absent in \cite{Sinamuli:2016rms}, more details can be found in appendix \ref{geonbulk}. The result is consistent with the CFT computation.

\section{Massless BTZ black hole}\label{MBTZ}
In this section we consider a limit case($n\to\infty$) of the conical defect, which is the massless BTZ black hole. One can regard the conical defect as a spacetime which has a stationary particle at $r=0$. When the mass contribution of this particle approaches  a critical value it will become a black hole geometry with black hole mass $M=0$. The holographic duality of massless BTZ black hole has been studied for many years \cite{Strominger:1997eq,deBoer:2010ac,Balasubramanian:2005qu}. However, in the present paper we only consider its quantum information metric in dual gravity side and give a rough picture without any field theoretic detail.

To proceed, let us start with the ordinary non-rotating BTZ black hole (\ref{btzmetric}).
The quantum information metric of this black hole has been discussed in \cite{MIyaji:2015mia,Bak:2015jxd,Trivella:2016brw,Sinamuli:2016rms} and explicitly it is given by
\begin{equation}\label{btz qim}
G^{BTZ}_{\lambda\lambda}=\frac{2}{\pi \tilde{\kappa}^2}\left[\frac{\pi V_1}{8\epsilon_{BTZ}}
-\frac{\pi V_1}{2\beta}+\frac{2\pi^2 V_1}{\beta^2}\cdot\tau \cot\left(\frac{4\pi\tau}{\beta}\right)\right],
\end{equation}
where we have set AdS radius to one and $V_1=2\pi$. $\epsilon_{BTZ}$ stands the cutoff for the nonrotational BTZ. In order to match the two point function on the CFT with the bulk side we multiply a constant $2/\pi\kappa^2$ \cite{Trivella:2016brw}.

Then we notice that the massless BTZ black hole(MBTZ) can be obtained by taking a limit $M=J=0$ from the non-rotating BTZ black hole\eqref{btzmetric}. Explicitly, the metric of massless BTZ geometry is of the following form
\begin{equation}
ds^2 = -\frac{r^2}{L^2} \,dt^2
+ \frac{L^2}{r^2} dr^2 + r^2 d\theta^2.
\label{mbtzmetric}
\end{equation}
Above metric shows that the MBTZ has an interesting feature: its horizon coincides with the singularity, i.e. $r_+=0$. This feature may lead to many nontrivial properties: there is no Hawking radiation of this black hole, i.e., $T_H=0$; the free energy(energy) and heat capacity(entropy) are also zero: $F_{MBTZ}=-E_{MBTZ}=0,\ C_{MBTZ}=S_{MBTZ}=0$. Based on these properties, one naively think that the quantum information metric can be obtained by taking the limit $\beta\to \infty$ in (\ref{btz qim})
\begin{equation}\label{gbtz}
G^{BTZ}_{\lambda\lambda}(\beta\to\infty)=\frac{1}{4\kappa^2}\cdot \frac{V_1}{\epsilon'_{BTZ}},
\end{equation}
where we have used the cut-off $\epsilon'_{BTZ}=n\epsilon_{BTZ}$. So this remains a cut-off dependent quantity.

On the other hand, one can also obtain the massless BTZ geometry by taking the $n\to\infty$ limit of the conical defect metric (\ref{defmetric}). So the results in this paper also hold for the massless BTZ black hole. We take the limit $n\to\infty$ in the quantum information metric of the conical defect (\ref{2d df qim}), and the $\mathcal{O}(\epsilon'^k)(k\ge1)$ terms are suppressed by large $n$ and small $\epsilon'_{BTZ}$, then it yields 
\begin{equation}\label{gdefect}
G^{defect}_{\lambda\lambda}(n\to \infty)= n_2^{defect} \left(\frac{V_1}{\epsilon'_{BTZ}}-\frac{V_1}{n}\right)=\frac{n^2}{4\kappa^2}\left( \frac{V_1}{\epsilon'_{BTZ}}-\frac{V_1}{n}\right),
\end{equation}
where we have used $n^{defect}_{2}=n^2\cdot n^{AdS}_{2}$ and $n_2^{AdS}=1/4\kappa^2$ from (\ref{nd}).
\begin{figure}
\centering
\includegraphics[scale=0.4]{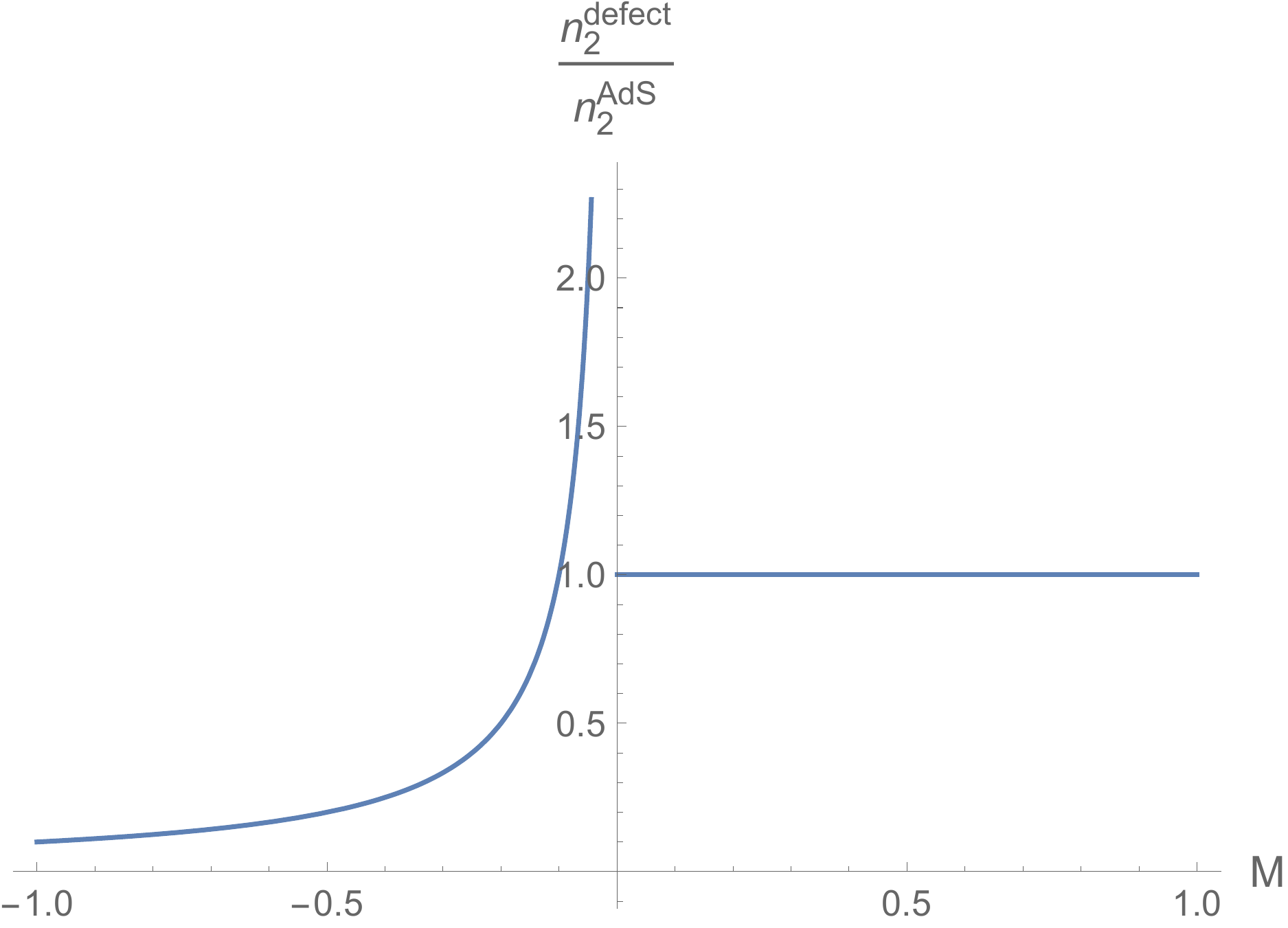}~~~~~~~~~
\includegraphics[scale=0.4]{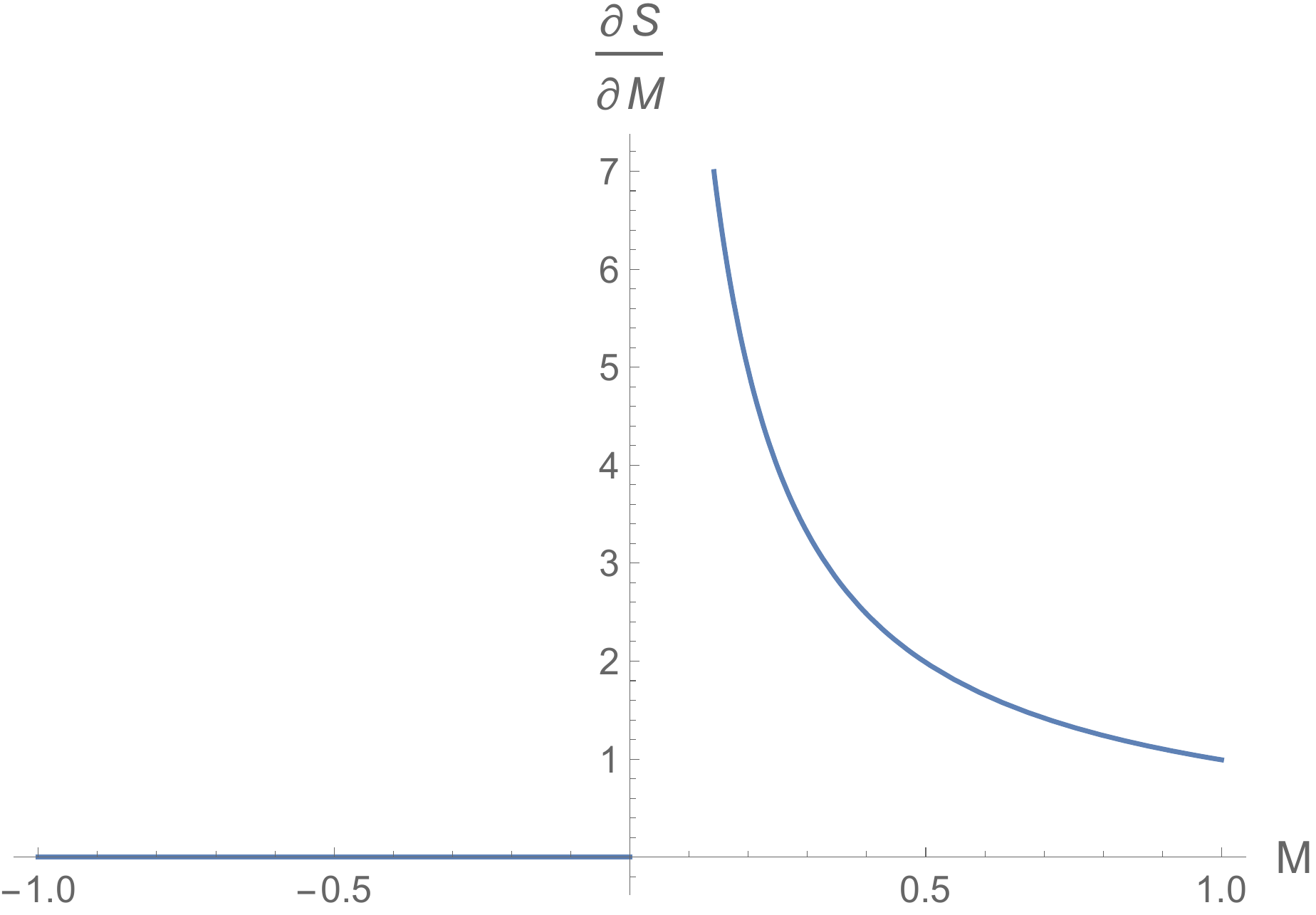}
\caption{We regard $M$ as an order parameter. (Left) The radio of the coefficient for the conical defect to the one for the AdS. There is a divergence when $M\to 0^-$. (Right) First derivative of $S$ with respect to $M$. There is a divergence when $M\to 0^+$. This indicates a second-order phase transition between BTZ and massless BTZ black hole. We set $8G=10$.}\label{figure}
\end{figure}
Eqs \eqref{gbtz} and \eqref{gdefect} indicate that in this limit the result dose not exactly coincides with $G^{BTZ}_{\lambda\lambda}(\beta\to\infty)$. Their cut-off dependent term have the similar behaviour but different coefficient: $n^{defect}_{2}=n^2\cdot n^{AdS}_{2}$. In addition, $G^{defect}_{\lambda\lambda}(n\to \infty)$ has a cut-off independent universal term. 

One natural question is the following:  which one is the correct quantum information metric of the massless BTZ black hole. We argue that the information metric of the massless BTZ is inclined to be obtained by taking $n\to\infty$ in the conical defect. The reason is the following.

Firstly, as we have explained above, the conical defect geometry can be constructed by exciting a stationary point particle at $r=0$. This particle in the interior also contributes to the whole spacetime. One can obtain the whole contribution by integrating the spacetime stress tensor of $\theta$ and give the global contribution \cite{Balasubramanian:1999zv,Balasubramanian:1999re}
\begin{equation}\label{adm mass}
M=-\frac{1}{8G}+\frac{\gamma}{8\pi G}.
\end{equation}
The first term in the r.h.s of \eqref{adm mass} corresponds to the pure AdS$_3$ contribution and the second term is the particle contribution, with $0\le \gamma \le \pi$. Physically, the range $0<\gamma<\pi$ corresponds to the total mass $-1/8G< M< 0$. When $\gamma=0$ this is the pure AdS case. The range $M\ge0$ corresponds to the spectrum of BTZ black holes (including massless BTZ black holes). In conical defect geometry the angular coordinate range is $\theta\in [0,2\pi A]$. We parametrize the conical defect as $A=1-4G\mu$, where $\mu$ is the mass of the particle \cite{Deser:1983nh}. This particle creates a singularity at $r=0$ which results in the conical defect metric with the deficit angle(FIG. \ref{scalarfield})
\begin{equation}\label{deficit angle}
\gamma=\pi(1-A)=4\pi G\mu.
\end{equation}
Note that the particle mass $\mu$ differs from the ADM mass contribution (\ref{adm mass}) and they are related by \cite{Myung:2005ee}
\begin{equation}\label{particle mass}
\mu=\frac{1}{4G}(1-\sqrt{-8GM}).
\end{equation}
The subalgebra of the conical defect geometry only has Virasoro form when $A=1/n$ \cite{deBoer:2010ac}, which leads to an AdS$_3/\mathbb{Z}_n$ as considered in this paper. Comparing (\ref{deficit angle}) and (\ref{particle mass}) we have
\begin{equation}
M=-1/8Gn^2.
\end{equation}
Clearly it becomes AdS as $n=1$ and massless BTZ black hole as $n=\infty$, as expected. This results in $n_2^{defect}/n_2^{AdS}=-1/8GM$, where $-1\le M\le 0$. So we find the coefficient of conical defect diverges when $M\to 0^-$(FIG. \ref{figure}).

One more evidence is that the non-rotating BTZ and massless BTZ are in different phases \cite{Myung:2006sq,Myung:2005ee,Cai:1996df}. The authors in \cite{Cai:1996df} provided some arguments that there is a second-order phase transition between the massless BTZ and non-rotating BTZ because the energy and heat capacity are continuous. Here we can also see this phase transition by calculating the behaviour of entropy in the bulk. To proceed, let us first focus on the free energy. The free energy of BTZ and conical defect was discussed in \cite{David:1999zb} and it can be obtained by calculating the on-shell action. To do this, firstly we need to regularize the divergence so it was suggested to consider the free energy of BTZ relative to AdS$_3$\cite{David:1999zb}

\begin{equation}
-\log Z=I(\text{BTZ})-I(\text{AdS})=\frac{1}{4\pi G}(\pi\beta-\pi^2r_+),
\end{equation}
where $\beta$ is the inverse temperature of BTZ.
The entropy of the BTZ black hole is then given by
\begin{equation}
S_{\text{BTZ}}=\left(1-\beta\frac{\partial}{\partial\beta}\right)\log Z_{\text{BTZ}}=\frac{\pi r+}{4G}.
\end{equation}
To discuss the massless BTZ black hole, we can choose the mass $M$ as an order parameter in the system. The entropy of the massless BTZ black hole is then obtained by taking the limit $\beta\to\infty$(or $M\to 0$)
\begin{eqnarray}
S_{\text{BTZ}}(\beta\to\infty)=0,
\end{eqnarray}
and the derivative w.r.t. $M$ is given by
\begin{eqnarray}
\frac{\partial S_{\text{BTZ}}}{\partial M}(\beta\to\infty)\sim\frac{1}{\sqrt{M}}\big|_{M\to 0}\to\infty.
\end{eqnarray}
We see that the derivative $\frac{\partial S_{\text{BTZ}}}{\partial M}$ becomes infinite when black hole tends to $M=0$.

On the other hand, the free energy of conical defect can be obtained very similarly and explicitly it is \cite{David:1999zb}

\begin{equation}
-\log Z=I(X_{\chi})-I(\text{MBTZ})=-\frac{\beta}{8G}\chi,~~~~~~(0\le\chi\le 1),
\end{equation}
where $X_{\chi}$ denotes the manifold of the conical defect and $\chi$ is related to deficit angle through $\gamma=\pi(1-\sqrt{\chi})$. From (\ref{particle mass}) we have $n^2=1/\chi$. In other words, the limit $n\to\infty$ implies $\chi\to 0$. We compute the free energy of conical defect relative to massless BTZ (which is also $X_0$) rather than AdS$_3$ because of the result matching with the partition function in dual CFT \cite{David:1999zb}. The free energy is proportional to $\beta$ so it has vanishing entropy for conical defect, $S_{X_{\chi}}=0$. The corresponding entropy and its derivative for $\chi\to 0$ (i.e., $M\to 0$ or $n\to \infty$) is
\begin{eqnarray}
&&S_{X_{\chi}}(\chi\to 0)=0,\\
&&\frac{\partial S_{X_{\chi}}}{\partial M}(\chi\to 0)=0.
\end{eqnarray}
It is obvious that from massless BTZ to massive BTZ there is a second order phase transition at $M=0$. As a contrast, there is nothing noncontinuous from conical defect to massless BTZ (FIG. \ref{figure})

As a last point, if we consider the quantum effect in the BTZ black hole, Lifschytz and Ortiz argued in \cite{Lifschytz:1993eb} that there is quantum instability of the massless BTZ and conclude that the end of Hawking evaporation is not the massless BTZ black hole but the AdS spacetime. So when the evaporation finish ($M= 0$) we may not simply relate to massless BTZ black hole in the bulk.

Based on all these observations, we conclude that one cannot simply make $\beta\to\infty$ in non-rotating BTZ black hole (\ref{btz qim}) to get properties for the massless BTZ black hole. Instead, it is more safe to obtain them from the conical defect by taking $n\to\infty$.

\section{Conclusion and Discussion}\label{sec:conclusion}

In this paper we have shown how to compute the quantum information metric of the conical defect dual theory not only in CFT side but also in holographic point of view. In CFT, the correlation function in conical defect dual theory can be constructed by summing over nontrivial correlation between different copy images, which are embeded in a larger space. The larger space is the covering space of the conical defect space, which is pure AdS$_3$. There are some nontrivial relations between the conical defect and its covering. One of the most important is that the Virasoro algebra of the conical defect dual theory is a subalgebra of that of its covering space. This leads to the fractionated spectrum and rescaled central charge $c=n\tilde{c}$. When we compute the quantum information metric of the conical defect dual theory, this nontrivial property results in a rescaled information metric comparing to the covering space: $G^{defect}_{\lambda \lambda}=n\cdot G^{AdS}_{\lambda \lambda}$. To verify our result, we perform a holographic computation, which is based on the method of \cite{Trivella:2016brw}. Specifically, we sum over all scalar field deformations of the action $\delta I$. It turns out that for marginal perturbations, only  $\delta I_2$ survives, and it can recover the result $G^{defect}_{\lambda \lambda}=n\cdot G^{AdS}_{\lambda \lambda}$.

We also consider the holographic description of the quantum information metric, which is the maximal volume of a $d$-dimensional spacelike surface in the bulk geometry (\ref{volume}). After comparing with our result, we find the dual relation for the conical defect is different from its covering by a coefficient, i.e., $n^{defect}_{2}=n^2\cdot n^{AdS}_{2}$. We argue that the difference comes from the $\mathbb{Z}_n$ quotient of the original space. This is further verified by the result in \cite{Sinamuli:2016rms}, where this coefficient for the geon black hole constructed by $\mathbb{Z}_2$ quotient of the original black hole is different from its covering by a factor $4$, that is, $n^{geon}_2 = 4\cdot n^{BH}_2$. This factor $4$ comes from the fact that the quantum information metric of a geon black hole is double of that of the original black hole and the volume of the geon black hole is half of that of the original one. We have a similar explanation for our conical defect result. In conical defect geometry, one copy of the conical defect is obtained by an identification $\theta\sim\theta+2\pi/n$ of the AdS$_3$ space. So the volume of the spacelike surface of the conical defect is $1/n$ of that of the AdS$_3$ space. Meanwhile, we have showed that the quantum information metrics of the conical defect and its covering are related by $G^{defect}_{\lambda \lambda}=n\cdot G^{AdS}_{\lambda \lambda}$. So similar to the geon example, this will give a factor $n^2$ between the coefficient in two theory.

As to the massless BTZ black hole, we find that the massless BTZ black hole and the massive BTZ are not in the same phase, instead, massless BTZ is in the same phase as the conical defect. This nontrivial property indicates that one cannot simply make $\beta\to\infty$ in non-rotating BTZ black hole (\ref{btz qim}) to get some properties of the massless BTZ black hole. The right way is to obtain them from the conical defect by taking $n\to\infty$.

\section*{Acknowledgements}
This work was supported in part by the National Natural Science Foundation of China under Grant Nos. 11465012 and 11665016, and the 555 talent project of Jiangxi Province.

\appendix
\section{Quantum information metric in holography}\label{review qim}
In this appendix we introduce the idea and some detailed computation for the quantum information metric in AdS$_{d+1}$, which mainly comes from \cite{Trivella:2016brw}. We first review the CFT set up and then give a holographic perspective.

One can consider a $d$-dimension CFT with Euclidean Lagrangian $\mathcal{L}_0$ and we add a deformation $\delta \mathcal{L}=\delta \lambda\mathcal{O}$ on it, where $\mathcal{O}$ is a primary operator with conformal dimension $\Delta$ and $\delta\lambda$ is a coupling constant of this deformation. The ground states of original theory and deformed theory are denoted by $\ket{\Psi_0}=\ket{\Psi(\lambda)}$ and $\ket{\Psi_1}=\ket{\Psi(\lambda+\delta \lambda)}$, respectively.

We use the path integral formalism to compute the overlap between the ground states of these two theories. First, for a generic state $\ket{\tilde{\varphi}}$ the overlap $\braket{\tilde{\varphi}|\Psi_0}$ can be written by
\begin{equation}
\braket{\tilde{\varphi}|\Psi_0}=\frac{1}{\sqrt{Z_0}}\int_{\varphi(\tau=0)=\tilde{\varphi}} \mathcal{D}\varphi \exp\left(-\int_{-\infty}^{0} d \tau \int d^{d-1}x \mathcal{L}_0 \right),
\end{equation}
where $Z_0$ is the partition function of the original theory:
\begin{equation}\label{z0}
Z_0=\int \mathcal{D} \varphi \exp\left(-\int_{-\infty}^{\infty} d \tau \int d^{d-1} x \mathcal{L}_0\right).
\end{equation}
In the same way, the overlap $\braket{\Psi_1|\tilde \varphi}$ can also be written in a similar formalism:
\begin{equation}
\braket{\Psi_1|\tilde \varphi}=\frac{1}{\sqrt{Z_1}}\int_{\varphi(\tau=0)=\tilde{\varphi}} \mathcal{D}\varphi \exp\left(-\int_{0}^{\infty} d \tau \int d^{d-1}x (\mathcal{L}_0+\delta \lambda \mathcal{O}) \right),
\end{equation}
where $Z_1$ is the partition function of the deformed theory:
\begin{equation}\label{z1}
Z_1=\int \mathcal{D} \varphi \exp\left(-\int_{-\infty}^{\infty} d \tau \int d^{d-1} x (\mathcal{L}_0+\delta \lambda \mathcal{O})\right).
\end{equation}
Then the overlap $\braket{\Psi_1|\Psi_0}$ can be obtained as
\begin{eqnarray}\label{overlap}
\braket{\Psi_1|\Psi_0}&=&\int \mathcal{D} \tilde \varphi \braket{\Psi_1|\tilde \varphi} \braket{\tilde \varphi|\Psi_0}\nonumber\\
&=&\frac{\int \mathcal{D} \varphi \exp \left(-\int_{-\infty}^{0} d \tau \int d^{d-1}x  \mathcal{L}_{0}- \int_{0}^{\infty} d \tau \int d^{d-1}x ( \mathcal{L}_0 +\delta \lambda \mathcal{O})\right)}{(Z_0 Z_{1})^{1/2}}. \label{overlap-pi}
\end{eqnarray}
It should be able to note that there is a discontinuity at $\tau=0$, which will result in UV divergences. So we introduced the UV regularization $\epsilon$ at $\tau=0$ to make the path integral well-defined. Then the $\ket{\Psi_1}$ is replaced by
\begin{equation} \label{reg state}
\ket{\Psi_1 (\epsilon)}=\frac{e^{-\epsilon H_0}\ket{\Psi_1}}{\left(\braket{\Psi_1|{e^{-2\epsilon H_0}|\Psi_1}} \right)^{1/2}},
\end{equation}
where $H_0$ is the Hamiltonian of the unperturbed theory. Now rewrite (\ref{overlap-pi}) as the expectation value formalism in the unperturbed ground state:
\begin{equation}\label{overlap2}
\braket{\Psi_1 (\epsilon)|\Psi_0}=\frac{\braket{\exp\left( -\int_{\epsilon}^{\infty} d \tau \int d^{d-1}x \delta \lambda \mathcal{O}(\tau, x)\right)}}{\braket{\exp\left( -(\int_{-\infty}^{-\epsilon}+\int_{\epsilon}^{\infty}) d \tau \int d^{d-1}x \delta \lambda \mathcal{O}(\tau, x)\right)}^{1/2}}.
\end{equation}
We can expand the absolute value of the overlap (\ref{overlap2}) for small $\delta\lambda$ to the second order, then we obtain
\begin{equation}
|\braket{\Psi_1 (\epsilon)|\Psi_0}|=1-G_{\lambda \lambda} \delta \lambda^2+\mathcal{O}(\delta \lambda^3),
\end{equation}
The quantum information metric is defined as the coefficient of the second-order term $G_{\lambda \lambda}$. From (\ref{overlap2}) we have
\begin{equation}
G_{\lambda \lambda} =\frac{1}{2}\int d^{d-1}x_1 \int d^{d-1}x_2\int_{-\infty}^{-\epsilon} d \tau_1 \int_{\epsilon}^{\infty}d \tau_2\braket{\mathcal{O}(\tau_1,x_1)\mathcal{O}(\tau_2,x_2)}
\end{equation}
where we have assumed the one-point function equal to zero, i.e. $\braket{\mathcal{O}}=0$ for operator in the original theory and using the time reversal symmetry relation $\braket{\mathcal{O}(-\tau_1,x_1)\mathcal{O}(-\tau_2,x_2)}=\braket{\mathcal{O}(\tau_1,x_1)\mathcal{O}(\tau_2,x_2)}$.
We can also define a partition function $Z_2$ which is the theory only deformed for $\tau>0$:
\begin{equation}
Z_2=\int \mathcal{D} \varphi \exp \left(-\int_{-\infty}^{0} d \tau \int d^{d-1}x  \mathcal{L}_{0}- \int_{0}^{\infty} d \tau \int d^{d-1}x ( \mathcal{L}_0 +\delta \lambda \mathcal{O})\right).
\end{equation}
Then the overlap (\ref{overlap}) can be written as
\begin{equation}
\braket{\Psi_1|\Psi_0}=\frac{Z_2}{(Z_1 Z_{0})^{1/2}}.
\end{equation}
\begin{figure}
\centering
\includegraphics[scale=0.8]{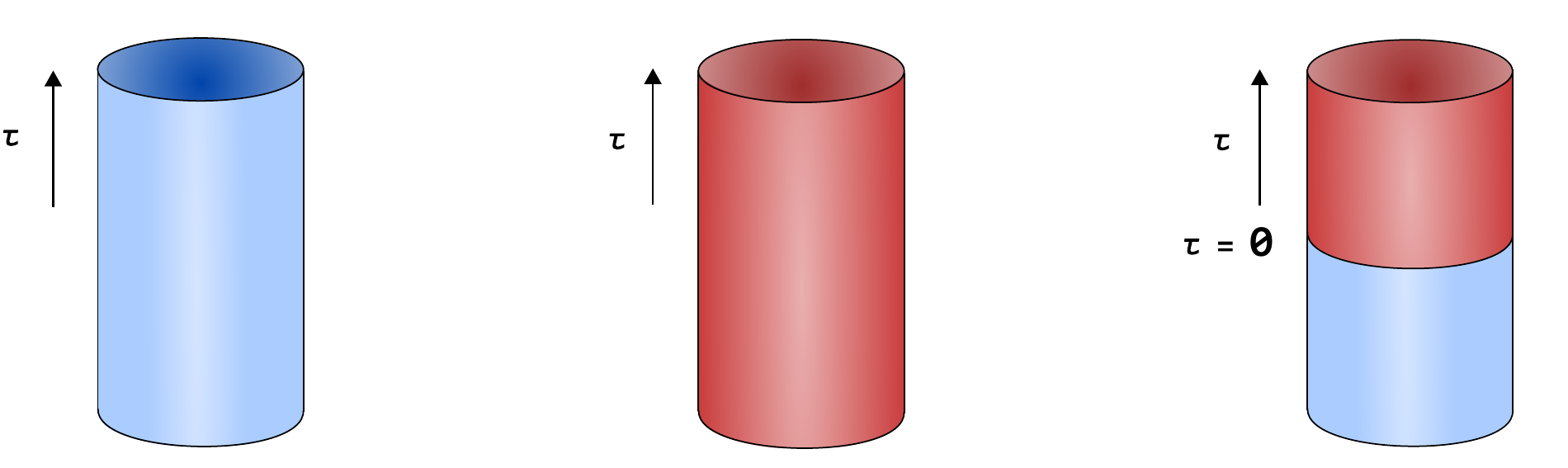}
\caption{A figure representation of configurations corresponding to $Z_0$, $Z_1$ and $Z_2$ on a cylinder. The blue color represents the undeformed theory and the red color represents the deformed theory induced by adding $\int \delta\tilde{\lambda}\tilde{\mathcal{O}}$ to the action.}\label{z0z1z2}
\end{figure}
This is the general definition of quantum information metric in CFT side. We can also represent it in the gravity side.

The partition function of field theory can be obtained from the gravity partition function in the classical limit according to the AdS/CFT conjecture. In this classical regime, only saddle points contribute to the partition function, i.e. $Z_k=e^{-I_k}$, where $k=0, 1, 2$ and $I_k$ is the on-shell action in the corresponding bulk.

Now let us turn to the bulk computation. From the holographic point of view, a probe  scalar field in the bulk corresponds to a deformation by operator in the boundary theory. So in the bulk we should consider a scalar-field perturbation to the unperturbed on-shell action, i.e. $I_k=I_{AdS}+\delta I_k$. The perturbation is given by
\begin{eqnarray}
\delta I_k&=&\frac{1}{2\kappa^2}\int d^{d+1}x \sqrt{g_0} \left(g^{\mu \nu}_0\partial_\mu \Phi_k \partial_\nu \Phi_k +m^2\Phi_k^2  \right)\\
&=&\frac{1}{2 \kappa^2}\int_{\partial \mathcal{M}}\sqrt{\gamma_0} n_{\mu} g^{\mu \nu}_0 \Phi_k \partial_\nu \Phi_k,
\end{eqnarray}
where $\Phi_1$ and $\Phi_2$ satisfy different boundary conditions. Then we can represent the overlap as
\begin{eqnarray}
\braket{\Psi_1|\Psi_0}&=&\frac{Z_{2}}{\sqrt{Z_1 Z_0}}=\exp \left(-I_{AdS}-\delta I_{2} + \frac{1}{2}(I_{AdS}+\delta I_{1}+I_{AdS})\right) \nonumber\\
&=& \exp \left( -\delta I_{2} + \frac{1}{2} \delta I_{1} \right).
\end{eqnarray}
The scalar field that obeys the Laplace equation in the bulk has leading term from non-normalizable mode near the boundary. In other words, the boundary condition of this field is given by
\begin{equation}
\lim_{r \to \infty} r^{d-\Delta} \Phi (r, \tau, \Omega)=\delta \lambda s_k(\tau),
\end{equation}
where
\begin{equation}
s_0(\tau)=0\nonumber ,~~~~~
s_1(\tau)=1 \nonumber ,~~~~~
s_2(\tau)=\begin{cases}
1&\text{if}\phantom{aa}\tau\ge 0\\
0&\text{if}\phantom{aa}\tau\leq 0.
\end{cases}
\end{equation}
The subscript indicates that for different $Z_k$ we chose different boundary condition of the scalar field. We focus on the marginal deformation $\Delta=d$ which is only relevant to $\delta I_2$. Under the change of coordinates
\begin{equation}
r=\frac{\cosh\rho}{\sinh T},~~~~~~~\tau=\frac{1}{2}\log\frac{\cosh(T+\rho)}{\cosh(T-\rho)},
\end{equation}
with $T>0, \rho\in\mathbb{R}$, the global AdS metric change to
\begin{equation}
ds^2=d\rho^2+\left( \frac{\cosh\rho}{\sin T}\right)^2(dT^2+g_{\mathbb{S}^d-1}).
\end{equation}
Together with the boundary condition, one can solve the equation of motion for $\Phi_2$, the solution is given by
\begin{equation}
\Phi_2(\rho)=\frac{\delta\lambda\Gamma(\frac{1+d}{2})}{\sqrt{\pi}\gamma(d/2)}\int_{-\infty}^{\rho}\frac{1}{\cosh^d x}dx.
\end{equation}
$\Phi_2$ only depends on $\rho$. By substituting $\Phi_2$ into the on-shell action, one obtains
\begin{equation}
\delta I_2=\frac{L^{d-1}\delta\lambda^2 \Gamma\left(\frac{1+d}{2}\right)}{2 \kappa^2 \sqrt{\pi} \Gamma(d/2)} \text{Vol}(\mathbb S^{d-1})\int_{\text{cut-off}}^{\infty}\frac{1}{\sinh^d T}dT,
\end{equation}
where we have put a cut-off at $\sinh T=\epsilon$. After changing $r=(\sinh T)^{-1}$, the following quantum information metric is obtained
\begin{equation}\label{globalG}
G^{AdS}_{\lambda \lambda}=\frac{L^{d-1} \Gamma\left(\frac{1+d}{2}\right)}{2 \kappa^2 \sqrt{\pi} \Gamma(d/2)} \text{Vol}(\mathbb S^{d-1}) \int_{0}^{1/\epsilon} \frac{r^{d-1}}{\sqrt{1+r^2}}dr.
\end{equation}
In \cite{MIyaji:2015mia} this prescription in gravity side is simply by adding the defect brane action $S_{brane}=T\int_{\Sigma}\sqrt{g}$ to the Einstein-Hilbert action. When the deformation is infinitesimally small the brane tension is $T\simeq n_d^{AdS}\delta \lambda^2/L^d$, where $n_d$ is fixed by
\begin{equation}\label{nd}
n^{AdS}_d=\frac{L^{d-1}  \Gamma\left(\frac{1+d}{2}\right)}{2 \kappa^2 \sqrt{\pi} \Gamma(d/2)}.
\end{equation}
which is the coefficient in (\ref{globalG}). This is a universal constant in the AdS$_{d+1}$ case.

So far we have derived the quantum information metric in Poincare AdS$_{d+1}$. In this paper, we consider the global AdS$_3$ case with its dual CFT living on a cylinder. The derivation of the quantum information metric in this situation is similar to the Poincare case. The quantum information reads \cite{Bak:2017rpp}
\begin{equation}
G_{\lambda \lambda} =\frac{1}{2}\int d^{d-1}\Omega_1 \sqrt{g_{S^{d-1}}}\int d^{d-1}\Omega_2 \sqrt{g_{S^{d-1}}}\int_{-\infty}^{-\epsilon} d \tau_1 \int_{\epsilon}^{\infty}d \tau_2\braket{\mathcal{O}(\tau_1,\Omega_1)\mathcal{O}(\tau_2,\Omega_2)},
\end{equation}
where $Z_0$ is the partition function of the unperturbed theory, $Z_1$ is the partition function of the perturbed theory for $-\infty<\tau<\infty$ and $Z_2$ is the partition function of the theory which is only perturbed for $\tau>0$, as shown in FIG. \ref{z0z1z2}.  $\Omega$ is the compact spacelike dimension $\mathbb{S}^{d-1}$ and $\tau$ is the noncompact timelike dimension. 

\section{Normalization constant of the two-point function}\label{normalizationconstant}
The derivation of the normalization constant of the two-point function (\ref{normalization}) is got by following \cite{Bak:2017rpp}. The bulk-to-boundary propagator in Poincare coordinates is 
\begin{equation}
K(x;x',z)=c_{\Delta} \frac{z^\Delta}{(z^2+|x-x'|^2)^{\Delta}}.
\end{equation}
The coefficient $c_{\Delta}$ can be fixed by reaching to the boundary $z\to 0$
\begin{equation}
K(x;x',z)=z^{d-\Delta}\delta^d(x-x').
\end{equation}
Thus we have
\begin{equation}
1=\int d^dx z^{\Delta-d} K(x;x',z)=c_{\Delta}\int d^dx \frac{z^{2 \Delta-d}}{(z^2+|x-x'|^2)^{\Delta}}.
\end{equation}
Algebraic calculation of the integral on the right hand side yields
\begin{equation}
c_{\Delta}=\left(\mathrm{Vol}{(S^{d-1})}\frac{\Gamma \left(\frac{d}{2}\right) \Gamma \left(\Delta -\frac{d}{2}\right)}{2 \Gamma (\Delta )}\right)^{-1}.
\end{equation}
Using this bulk-to-boundary propagator the bulk scalar field can be obtained by
\begin{equation}
\phi(x,z)=c_\Delta \int d^dx' \frac{z^\Delta J(x')}{(z^2+|x-x'|^2)^{\Delta}}.
\end{equation}
where $J(x)$ is the boundary condition of $\phi(x,z)$. The normalization constant $\mathcal{N}_{\Delta, \kappa}$ defined in \eqref{normalization} has to be fixed from the holographic point of view: $\braket{\exp({\int J \mathcal{O}})}_{\mathrm{CFT}}=\exp{(-I)}$, where $I$ is the on-shell action of the bulk theory
\begin{equation}
I=-\frac{1}{2\kappa^2} \int_{z=\epsilon} d^{d}x (\sqrt{g}g^{zz}\phi \partial_z \phi ).
\end{equation}
Finally we have $\mathcal{N}_{\Delta, \kappa}=\ell^{d-1} c_\Delta d/\kappa^2$. The CFT on $\mathbb{R}\times S^{d-1}$ is related to the Poincare case by a conformal map from plane to cylinder. So the two-point function in CFT on a cylinder has the same constant $\mathcal{N}_{\Delta, \kappa}$.

\section{The bulk computation of QIM of $\mathbb{RP}_2$ geons}\label{geonbulk}
Similar to the conical defect case, we add a marginal perturbation in the ground state and see the change of the bulk action $\delta I_2$. In our argument this includes contributions from scalar fields in two images:
\begin{equation}
\delta I_2^{\text{geon}}\sim \int \sqrt{\gamma_0} n_{\mu} g^{\mu \nu}_0 \Phi_2(x) \partial_\nu \Phi_2(x)+\int \sqrt{\gamma_0} n_{\mu} g^{\mu \nu}_0 \Phi_2(\tilde{J}x) \partial_\nu \Phi_2(\tilde{J}x).  \label{geonbulkcontribution}
\end{equation}
To obtain $\delta I_2^{\text{geon}}$ we work in the unperturbed black hole metric, which can be written as
\begin{equation}
ds^2=\frac{dZ^2}{(1-Z^2)Z^2}+\frac{1-Z^2}{Z^2}d\hat{\tau}^2+\frac{d\phi^2}{Z^2},
\end{equation}
where the following coordinate transformation has been performed
\begin{equation}
x=\sqrt{1-Z^2}\cos \hat{\tau}e^{\phi}~~~~~~~~y=\sqrt{1-Z^2}\sin \hat{\tau}e^{\phi}~~~~~~~~z=Ze^{\phi}.
\end{equation}
To proceed, we first see the contribution from the first term, which is just the original BTZ. By introducing a cutoff at $Z=\epsilon e^{-\phi}$ we have \cite{Trivella:2016brw}
\begin{equation}\label{perturbation in BTZ}
\delta I_2^{\text{First}}\sim -\int d\phi d\hat{\tau}\frac{e^{2\phi}}{\epsilon}\Phi_2\partial_z\Phi_2\big{|}_{z=\epsilon},
\end{equation}
where we have set $r_+=1$ ($\beta=2\pi$) for simplicity. The $\hat{\tau}$ integral will run over $\hat{\tau}\in[-\pi+\tau, -\tau]$. The scalar field $\Phi_2$ has the boundary condition similar to (\ref{coupling const}) with no vanishing value in $\hat{\tau}\in[-\pi+\tau, -\tau]$. Since the integral is even in $\hat{\tau}$ we can only limit to $\hat{\tau}\in[-\pi/2, -\tau]$ and then multiply a factor $2$ in front of the integral. Then we shift variable of integration $\hat{\tau}'=\hat{\tau}-\epsilon$ and take the $\epsilon$ limit, then we obtain \cite{Trivella:2016brw}:
\begin{equation}
\partial_z \Phi(\hat{\tau}'+\epsilon, \phi)\big{|}_{z=\epsilon}=\frac{\delta\lambda}{\pi}e^{-2\phi}\mathcal{F}(\hat{\tau}')\epsilon,
\end{equation}
where
\begin{eqnarray}
&&\mathcal{F}(\hat{\tau})=\frac{\mathcal{F}_1(\hat{\tau})}{2(\cos(2T)-\cos(2\hat{\tau}))^2},\\
\nonumber\\
&&\mathcal{F}_1(\hat{\tau})=2\cos(2\hat{\tau})[\sin(2T)+\pi\cos(2T)]+h_1(\hat{\tau})+h_1(-\hat{\tau})-\sin(4T)-2\pi,
\end{eqnarray}
and $h_1$ is a function of $\hat{\tau}$. Finally, by integrating the $\phi$ in $\delta I_2$ we have
\begin{equation}\label{geon bulk}
\delta I_2^{\text{First}}\sim -\frac{\delta\lambda^2 V_1}{\pi}\int_{-\pi/2-\epsilon}^{-\tau-\epsilon} d\hat{\tau}'\mathcal{F}(\hat{\tau}')=\frac{\delta\lambda^2 V_1}{\pi}\left(\frac{\pi}{2\epsilon}+\frac{-1+2\tau \cot(2\tau)}{2}\right).
\end{equation}

Now we turn to consider the second contribution in (\ref{geonbulkcontribution}). After a changing of coordinates $Z=r_+/r$, $r_+\hat{\tau}=it$ and $\phi=r_+\theta$ , we get the Lorentzian metric (\ref{btzmetric}) by setting $r_+=1$. So the second contribution comes from the quotient $(\hat{\tau}, \phi)\sim(-\hat{\tau}, \phi+\pi)$. We can simply replace $(\hat{\tau}, \phi)$ with $(-\hat{\tau}, \phi+\pi)$ in the integral (\ref{perturbation in BTZ}) and obtain
\begin{equation}
\delta I_2^{\text{Second}}\sim -\frac{\delta\lambda^2 V_1}{\pi}\int_{-\pi/2-\epsilon}^{-\tau-\epsilon} d\hat{\tau}'\mathcal{F}(-\hat{\tau}')
\end{equation}
We note that this integral is independent of $\phi$ and $\mathcal{F}(\hat{\tau})$ is even in $\hat{\tau}$, i.e. $\mathcal{F}(-\hat{\tau})=\mathcal{F}(\hat{\tau})$. So the second contribution is the same as the first one i.e. $\delta I_2^{\text{Second}}=\delta I_2^{\text{First}}$. Finally the quantum information metric of geon black hole is double of one of the original black hole, which agrees with the CFT computation.

\end{document}